\documentclass[a4paper]{jpconf}
\usepackage{graphicx}
\usepackage{subfigure}
\usepackage{multirow}
\usepackage{isotope}

\begin{document}
\title{Efficiency calibration of the BRIKEN detector: the world largest counter for beta-delayed neutrons}

\author{M. Pall\`as$^1$, A. Tarife\~no-Saldivia$^1$, F. Calvi\~no$^1$, N. Mont-Geli$^1$, J. L. Tain$^2$, A. Tolosa-Delgado$^2$, J. Agramunt$^2$, F. Molina$^3$, P. Aguilera$^3$, J. Romero-Barrientos$^3$}

\address{$^1$Institut de Tècniques Energ\`etiques (INTE), Universitat Polit\`ecnica de Catalunya (UPC), Barcelona, Espa\~na.\\
$^2$Instituto de F\'isica Corpuscular (IFIC), Val\`encia, Espa\~na.\\
$^3$Departamento de Ciencias Nucleares, Comisión Chilena de Energ\'ia Nuclear, Santiago, Chile.  }
\ead{max.pallas@upc.edu, ariel.esteban.tarifeno@upc.edu }
\begin{abstract}
$\beta$-delayed neutron emission plays a fundamental role in the explosive nucleosynthesis of elements heavier than iron by the rapid neutron capture (r-process). The most ambitious project related to $\beta$-delayed neutron detection of very exotic nuclei is carried out by the BRIKEN collaboration at RIKEN Nishima Center. In this work, a brief description of the BRIKEN project is presented. A methodology for the precise characterization of the BRIKEN neutron counter efficiency, for fast neutrons, using an uncalibrated \isotope[252]{Cf} neutron source is described. The method relies on the well-known neutron multiplicity distribution of such source and correlation counting method. A detailed experimental study with the BRIKEN neutron counter and a \isotope[252]{Cf} neutron source at the RIKEN Nishina center is presented. The result of this work is the accurate determination of the neutron detection efficiency of the BRIKEN neutron counter with high accuracy (relative uncertainty $\sim 0.5\%$).
\end{abstract}
\section{Introduction}
$\beta$-delayed neutron emission is an event that rarely happens by natural means on earth but is a dominant decay process on very neutron-rich nuclei. This decay mode is key to describe the explosive nucleosynthesis of elements heavier than iron by the rapid neutron capture (r-process). At this moment, one of the goals of nuclear astrophysics research is to determine the nuclear properties of these neutron-rich nuclei. In particular, the neutron-emission branching ratios ($P_{xn}$) and the $\beta$-decay half-lives ($T_{1/2}$). The main challenge for this kind of research is the production of isotopes far from the valley of $\beta$-stability (radioactive beams) and to measure accurately these nuclear properties. The BRIKEN project aims to expand our current knowledge on $P_{xn}$ and $T_{1/2}$ values for the most exotic currently accessible neutron-rich nuclei. To achieve its aim, the BRIKEN project exploits the very intense radioactive beam by in-fight production from the Radioactive Isotope Beam Factory (RIBF) \cite{okunoetal2012}, and state-of-the-art nuclear instrumentation: (i) the AIDA implanted ion and decay detector \cite{griffinetal2017}, (ii) the BRIKEN neutron counter, the world largest $\beta$-delayed neutron counter \cite{conceptual}. When using moderated neutron counters and implantation detectors, the $P_n$-value for the decay of a specific nucleus is determined from the counting of beta-decays $(N_\beta)$, $\beta$-delayed neutrons $(N_n)$ and the detection efficiencies. Thus, for the single $\beta$-delayed neutron emission of a particular nucleus, the $P_n$-value can be written as \cite{agramuntetal2015_characterization},
\begin{equation}
    P_{1n}=\frac{\bar{\epsilon}_\beta}{\bar{\epsilon}_n}\frac{N_n}{N_\beta},
\end{equation}
Where $\bar{\epsilon}_\beta$ and $\bar{\epsilon}_n$ are the beta and neutron detection efficiencies, respectively, averaged over the possible $\beta$ and neutron energies of the decay. The last equation illustrates the importance of the proper characterization of the detection efficiencies, in particular the neutron efficiency, for the determination of the decay properties. In this work is presented the first systematic study of the BRIKEN neutron detection efficiency. 

\section{Experimental set up}
\label{NeutronDetector}
\begin{figure}[ht]
    \centering
    \includegraphics[scale=1.]{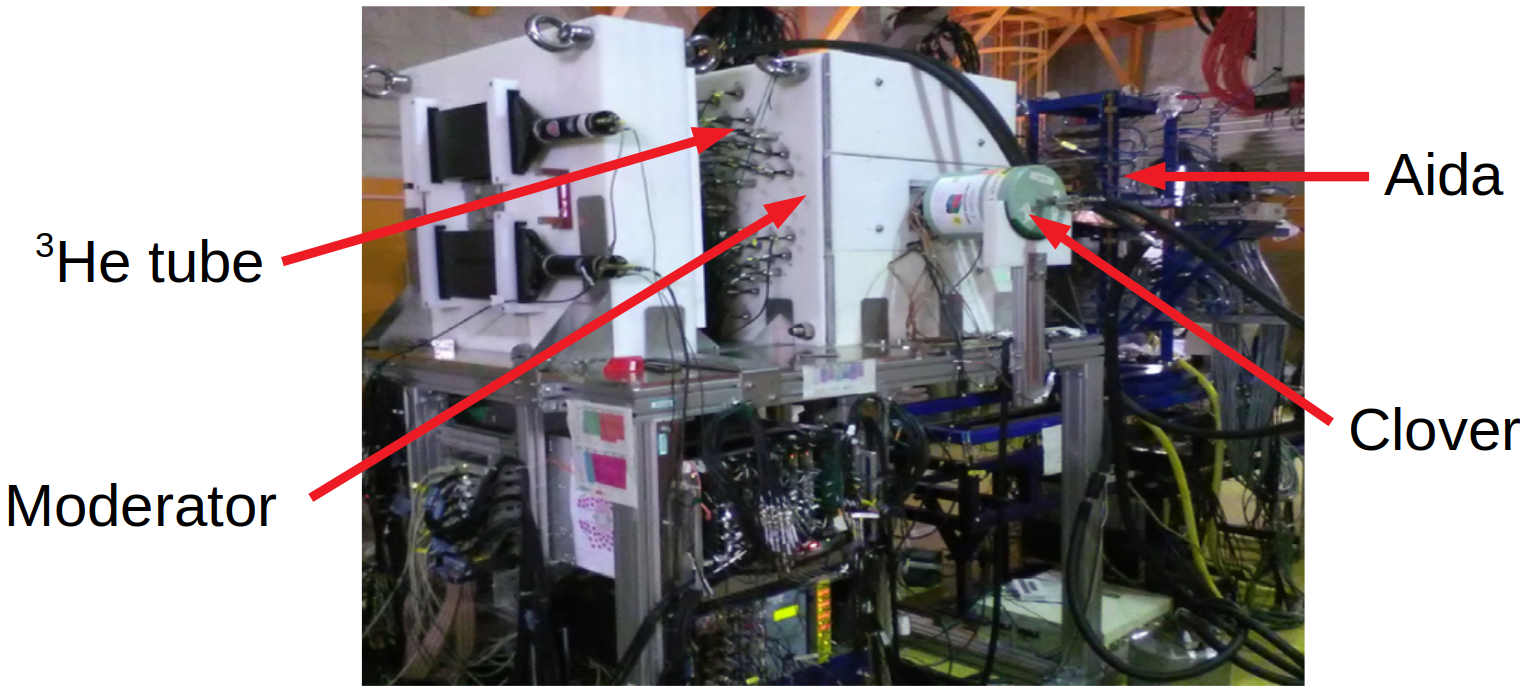}
    \caption{Experimental setup for the BRIKEN project.}
    \label{Setup} 
\end{figure}

The detector was designed in early 2016 to achieve the best possible performance in terms of neutron detection with a flat response up to $1\,\mathrm{MeV}$ \cite{conceptual}. The moderator is a HDPE block and has dimensions of $90\times90\times75\,cm^3$. On both sides of the block and perpendicularly to the beam, there are holes where clover type HPGe $\gamma$ detectors can be placed (Figure \ref{Setup}). The distribution of the tubes is symmetric in the plane perpendicular to the beam and follows an approximate ring geometry. Another peculiarity of this detector is the use of a digital data acquisition system (DAQ) that provides an absolute timestamp for every registered event \cite{commissioning}.
\\
For the detector characterization, an uncalibrated \isotope[252]{Cf} source has been used. This means that we do not have information about the exact number of neutrons emitted as a function of time. The data sheet handled by the manufacturer offered a nominal activity of $3.7\,\mathrm{MBq}$ and the impurities.

\section{Theoretical framework for data analysis}
\label{DataAnalysis}
\begin{figure}[ht]
    \centering
    \includegraphics[scale=1.2]{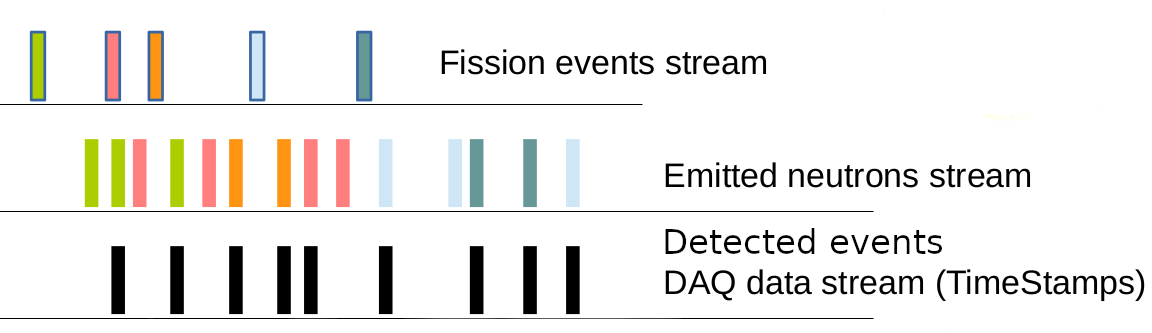}
    \caption{Scheme of the neutron detection from a spontaneous fission event. Each fission event can emit several neutrons. The neutrons emitted from the same fission event are correlated (same color). Some of them, depending on the efficiency, are detected by the neutron counter, from \cite{arielpower}.} 
     \label{DataStream}
\end{figure}

\begin{figure}[ht]
    \centering
    \includegraphics[scale=.5]{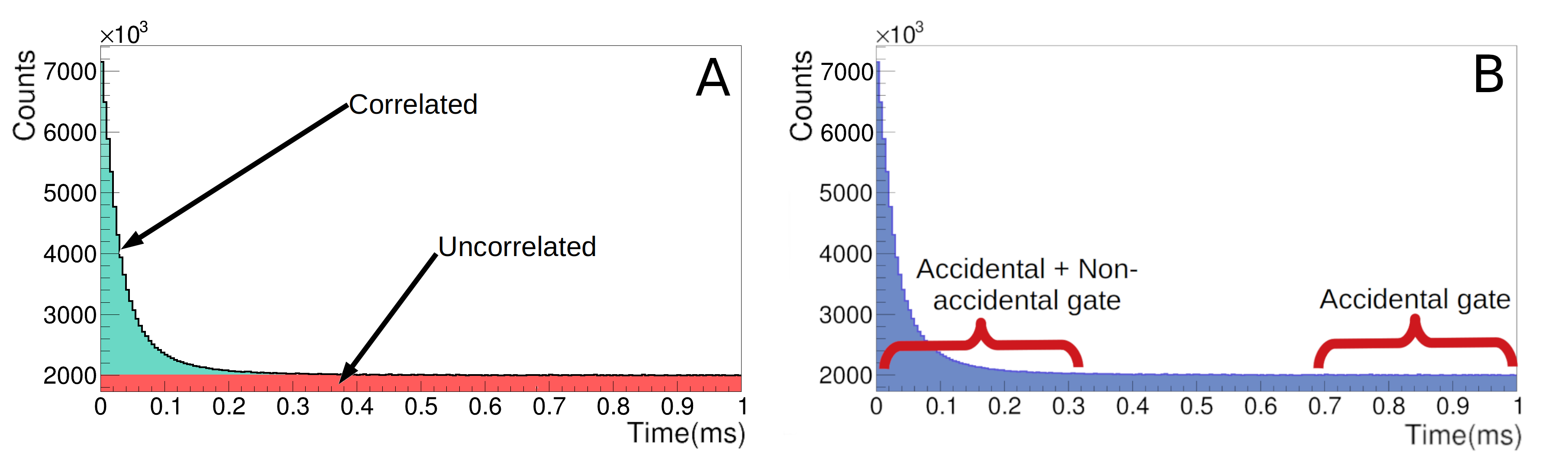}
    \caption{Figure A shows the two regions in a Rossi-$\alpha$ distribution. The green one is produced by fission neutrons correlated to the initial signal. The area of this region is the number of double neutron events. The red area is produced for the accidental correlation of uncorrelated events. Notice that these correlations are random. For this reason, this region is on the average constant. Figure B illustrates the implementation of the gate method. The first gate is opened in the region where we have a combination of accidental and non-accidental coincidences. While the second is opened in a time where our distribution only includes accidental coincidences.}
     \label{RossiAlpha}
\end{figure}

The efficiency of a detector can be determined directly by comparing the emission rate of the source with the detected rate. Unfortunately, we had available an uncalibrated \isotope[252]{Cf} source. For this reason, is necessary to develop an alternative method to determine the efficiency. In the BRIKEN neutron counter case, the methodology proposed for this study is based on Neutron Multiplicity Counting (NMC). Despite the source is not calibrated, the NMC can provide an accuracy lower than $1\%$ in the determination of the total neutron yield \cite{henzlova}. Neutron multiplicity counting (NMC) is an experimental technique used for non-destructive assay of special nuclear materials \cite{alamos}. This technique is often based on the use of moderated \isotope[3]{He} counters and coincidences counting electronics. Figure \ref{DataStream} illustrates the detection process. The \isotope[252]{Cf} suffer from spontaneous fission. Each of these fission can produce several neutrons that are time-correlated. Using the \isotope[3]{He} tubes we detect some of the emitted neutrons with an associated timestamp. This neutron pulse stream contains both correlated and uncorrelated events. The detection of two neutrons that were produced in the same fission is called a double neutron event.
The number of double neutron event $(D)$ is key to calculate the detector efficiency for a \isotope[252]{Cf} source as \cite{alamos}:
\begin{equation}
	\epsilon = \frac{2\nu_{s1}}{\nu_{s2}}\frac{D}{S}
	\label{Eq:Efficiency}
\end{equation}
Where $S$ is the number of individual events detected and $\nu_{si}$ are the reduced moments of the spontaneous fission neutron distribution. Using the efficiency we can calculate the neutron yield as: $Y=S/\epsilon$.
\\
Using a data stream of detected events ordered according to the timestamps, we can separate the correlated neutron events from the background of uncorrelated neutron events using the so-called Rossi-$\alpha$ distribution. It is the distribution in time of events that follow after an arbitrarily chosen starting event. If all the events are uncorrelated, the distribution is constant in time. On the other hand, if the signals present a correlation, the plot shows an exponential decay which is related to the moderation plus capture the of the neutrons (Figure \ref{RossiAlpha}). The area of the histogram above the flat background provides the rate of double neutron events detected.
We will calculate the rate of double neutron events using two different but equivalent methodologies.
\subsection{Rossi-$\alpha$ method}
The first method consists of the direct integration of the area of double coincidence events subtracting the extrapolated linear background fitted in the region of very large correlation times. We called this method the Rossi-$\alpha$ method (Figure \ref{RossiAlpha}A).
\subsection{Gate method}
The second option is what we called the gate method. The idea of the gate method is to extract the multiplicity distribution in two different time windows of in our correlation time data set. The first gate is opened in the region where we have a combination of accidental and non-accidental coincidences. The second is opened at a later time where our distribution only includes accidental coincidences (Figure \ref{RossiAlpha}B). Then, the neutron multiplicity distribution for each region is built and its first moment is calculated. Using these values we can prove that the rate of double neutron events is \cite{alamos}:
\begin{equation}
	D =  S(f_1-b_1)
\end{equation}
Where $S$ is the rate of singles (rate of total counts) and $f_1,\,b_1$ are the first moments of the accidental+non-accidental and accidental multiplicity distribution respectively.
\section{Experimental results and discussion}
\label{BRIKENefficiency}
\begin{figure}[ht]
    \centering
    \includegraphics[scale=1.4]{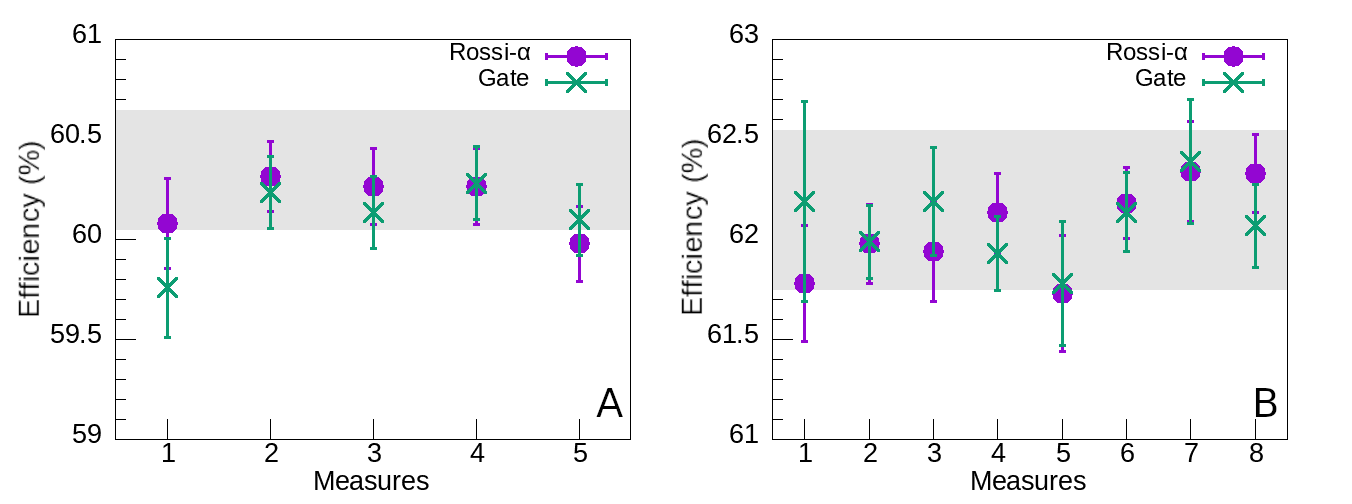}
    \caption{Efficiency for different measures. The right figure uses a setup without $\gamma$-detectors (clovers) while the left one includes clovers. Both methods give compatible values for both setups. The error bars were calculated propagating only the statistical uncertainties. The grey area is the efficiency recommendation for each setup listed in Table \ref{Results}}
     \label{effiplot}
\end{figure}

\begin{figure}[ht]
    \centering
    \includegraphics[scale=0.75]{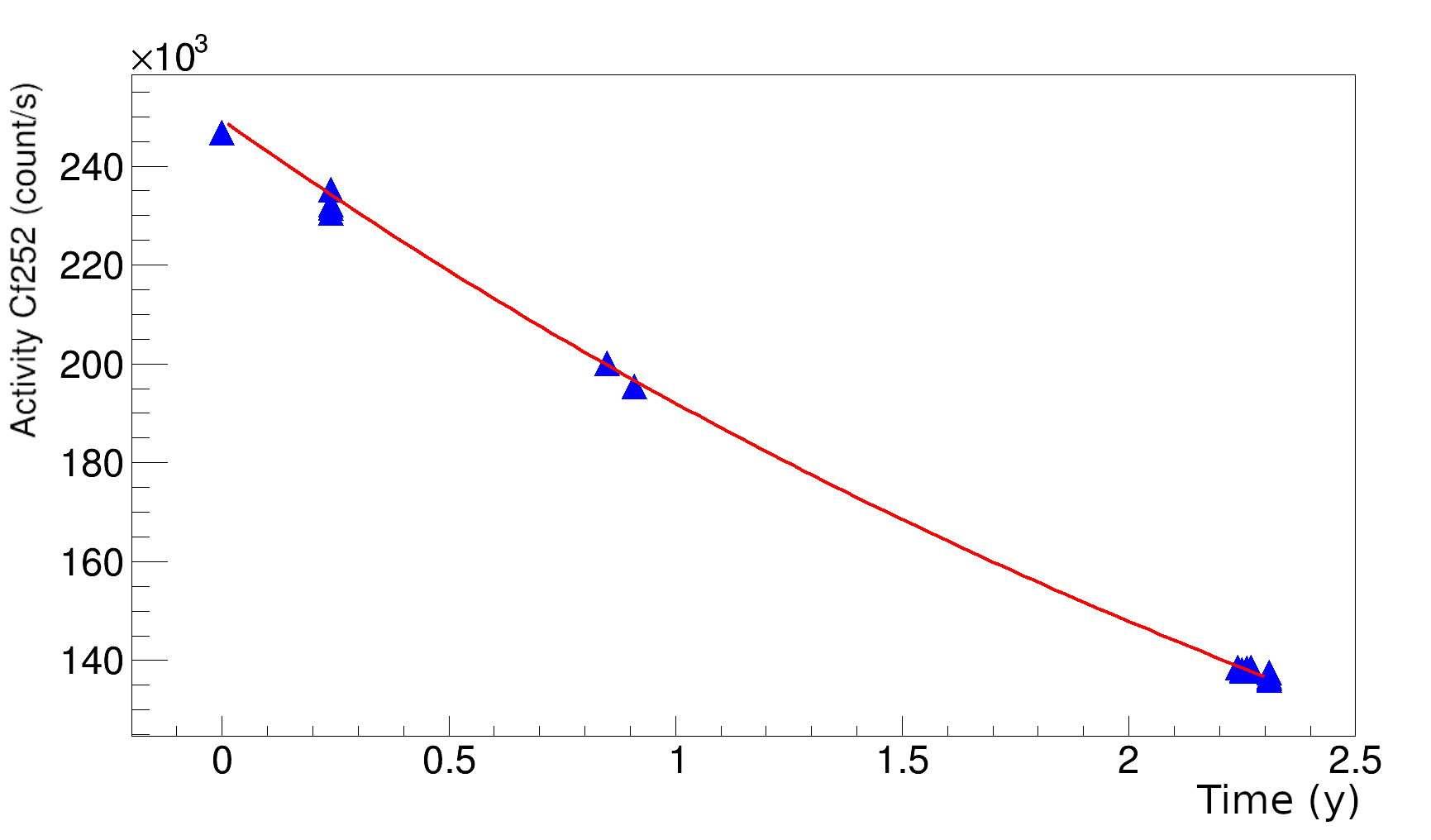}
    \caption{\isotope[252]{Cf} source activity measured. The markers are experimental data, error bars are lower than point size. The red line represents a fit to the points where the well-known \isotope[252]{Cf} half-life is fixed} 
     \label{activity}
\end{figure}

Once we have discussed the methodology we should discuss the experimental results for the neutron efficiency for the BRIKEN detector. To obtain a reliable value for the efficiency the experimental data include tens of measurements taken from 2016 to 2018. During this time some of the elements of the experimental setup were changed. In particular, the efficiency value changes considerably depending on the presence of the clovers, for that reason, we reported an efficiency value for each case. Figure \ref{effiplot}A and \ref{effiplot}B show the efficiency calculated by the methods described above for a selection of measurements. We can conclude that both methods achieve consistent results for different measurements.
Averaging the results of different measurements we obtain efficiencies for the BRIKEN detector (Table \ref{Results}). For the estimated error we included the statistical correction plus some systematic sources of error such as the dead time. Due to the DAQ implementation, the dead time of this system can be described via the non-paralyzable approximation \cite{knoll}. This correction has to be applied tube per tube. Applying this model to our data we found that the correction is rather small ($\sim +0.05\%$). The other important source of systematic error is the presence of other nuclei in the \isotope[252]{Cf} source. We found the only one having a sizeable effect is \isotope[252]{Cf} ($\sim +0.1\%$) \cite{radeve}. Notice that both corrections are positive, which means that our error bands are asymmetrical.   
\begin{table}[ht]
\centering
\begin{tabular}{ccc}
\hline
Setup          & Efficiency (\%) & Error(\%)                                             \\ \hline
Without clover & 60.33           & \begin{tabular}[c]{@{}c@{}}+0.32\\ -028\end{tabular}  \\
With clover    & 62.14            & \begin{tabular}[c]{@{}c@{}}+0.41\\ -0.39\end{tabular} \\ \hline
\end{tabular}
\caption{Efficiencies calculated using both methods of analysis. The errors include the statistical propagation uncertainties and systematic errors such as the dead time or the isotopic contamination. The errors are asymmetrical due to the asymmetrical systematic errors.}
\label{Results}
\end{table}
Apart from achieving an efficiency value, we can use the measurements to calibrate the \isotope[252]{Cf} source. The total neutron emission rate ($Y(t)$) is expected to follow the exponential decay law:
\begin{equation}
    Y=Y_0\;e^{\lambda t}=Y_0\;e^{\frac{ln(2)\,t}{T_{1/2}}}
\end{equation}
Where $\lambda$ is the decay constant and $T_{1/2} = 2.645y$ is the half-life for \isotope[252]{Cf}. Using data taken over three years, we can evaluate the exponential decay of disintegration of the source. The results for the experimentally determined total emission rate of the neutron source are presented in Figure \ref{activity}. The relative uncertainties resulting from our analysis for the different measurements span from 0.5 \% to 0.6 \%.
\\

\section{Conclusions}
In this work is presented the first systematic study of the BRIKEN neutron detection efficiency. For this experiment, we used a \isotope[252]{Cf} uncalibrated source. To calculate the efficiency despite that, we implemented the NMC technique and discussed two equivalent ways to apply it. The results obtained for the efficiency of the BRIKEN neutron counter are presented in Table \ref{Results}. The setup without $\gamma$-detectors presents a lower efficiency $(\sim 60\%)$ than the one using clovers $(\sim 62\%)$ . Finally, using the same methodology we offered a precise calibration for the \isotope[252]{Cf} source.
\section{Acknowledgments}
This work has been supported by the Spanish Ministerio de Econom\'ia y Competitividad under grants:
FPA2011-24553, FPA2011-28770-C03-03, FPA2014-52823-C2-1/2. FPA2017-83946-C2-1/2, PID2019-104714GB-C21/2 and the program Severo Ochoa (SEV-2014-0398). FONDECYT Regular 1171467

\section*{References}
\bibliographystyle{unsrt}
\bibliography{Biblio}
\end{document}